\documentstyle[12pt] {article}
\begin{document}

\title{A Quantum-Mechanical Equivalent-Photon Spectrum for Heavy-Ion Physics}
\author{C.J. Benesh$^{1,2,3}$, A.C. Hayes$^4$, and J.L. Friar$^1$,\\
1-Los Alamos National Laboratory, Los Alamos, NM  87545\,\,\,\,\,\,\,\\
2-Department of Physics, University of Northern Iowa,\,\,\,\,\,\,\,\,\,\,\,\,\,\,\,\,\,\,
\\ Cedar Falls,
Ia. 50614\\ 
3-International Institute for Theoretical and Applied
Physics,\\Ames, Iowa 50011\\
4-AECL, Chalk River Laboratories,\,\,\,\,\,\,\,\,\,\,\,\,\,\,\,\,\,\,\,\,\,\,\,\,\,\,\,\,\,\,\,\,\,\,\,\,\,\,\,\,\,\qquad\qquad\qquad\qquad\qquad \\ Chalk River, Ontario, Canada, K0J 1J0 }

\maketitle

\begin{abstract}
 	In a previous paper, we calculated the fully quantum-mechanical 
cross section for electromagnetic excitation during peripheral heavy-ion
collisions. Here, we examine the sensitivity of that cross section to 
the detailed structure of the projectile and target nuclei. At the transition
energies relevant to nuclear physics, we find the cross section to be weakly 
dependent on the projectile charge radius, and to be sensitive to only the 
leading momentum-transfer dependence of the target transition form factors.
We exploit these facts to derive a quantum-mechanical ``equivalent-photon 
spectrum'' valid in the long-wavelength limit. This improved
spectrum includes the effects of
projectile size, the finite longitudinal momentum transfer required by 
kinematics, and the response of the target nucleus to the off-shell photon. 
\end{abstract}

\pagebreak

\section{Introduction}

	In the previous decade, relativistic heavy-ion beams have become a 
useful tool for the study of electromagnetic processes in nuclei. Applications
have included studies of nuclear astrophysics\cite{1}, nuclei far from 
stability\cite{2}, and searches for multi-phonon excitations in
nuclei\cite{3}. In these experiments, cross sections that are difficult
to measure by other means are amplified
by the projectile charge, and, in the case of relativistic projectiles, by the
contraction of the projectile's electric field into a sharp pulse. 

Almost 
exclusively, the data from these experiments have been analyzed using 
the semi-classical
Weizs\"acker-Williams method of virtual quanta\cite{4}, in which the
cross section for the heavy-ion-induced reaction
is calculated by integrating the cross section for the analogous real-photon
process over a flux of photons that is ``equivalent'' to those that make
up the electric field of the projectile. In its simplest form, the pulse
of equivalent photons is obtained from the boosted Coulomb field of the 
projectile by equating the classical Poynting flux onto the target to
the energy flux carried by the pulse of equivalent photons. The
semi-classical spectrum has been generalized to include arbitrary 
multipoles\cite{5}, projectile structure\cite{6} and 
Coulomb scattering effects\cite{7}. 
Attempts to move beyond the semi-classical picture of these processes have
been thwarted  by lack of information about the structure of the target 
nucleus\cite{8}.
Furthermore, there has been little motivation for improvement
because the semi-classical spectrum, when used  
in conjunction with data from real-photon 
processes, provides model-independent results for
cross sections measured in heavy-ion collisions\cite{9}. 

	Recently, we have undertaken a program to 
systematically examine corrections to the semi-classical picture\cite{10}
\cite{11},
and have found significant deviations from the predictions of the 
Weizs\"acker-Williams method for the mildly relativistic collisions
($\gamma < 2-3$) that constitute a significant fraction of the  
available 
data. The aim of the present work is to expand on the results of reference
10, examining the sensitivity of the cross section to nuclear structure
inputs. Having determined which inputs are essential to
extracting the correct physical cross sections, we construct a simple model
that incorporates these features, and is
valid in the limit of low transition energies.  
 This  allows us to obtain a new, fully quantum-mechanical
expression
for an 
``equivalent-photon spectrum''  that can be used with measured photoabsorption 
cross sections in exactly the same fashion as the semi-classical 
expression. 

	The paper is structured as follows: In the next section, 
we briefly review the results of reference 10, emphasizing the interplay 
between the different length scales that determine the cross sections 
measured in heavy-ion collisions.
The third section is devoted to a comparison of the exact numerical results
for the cross sections using an assortment of parametrizations for
the projectile form factors and target transition densities. In the fourth 
section, we use the results from these comparisons to construct a simple model
for the form factors and transition densities that incorporates the important
physical parameters of the projectile and target. This allows us to extract
a new effective photon spectrum that may be used in the same fashion as the
Weizs\"acker-Williams spectrum. In the last section, we compare the predictions
obtained using the new spectrum with selected data from heavy-ion collisions.  
 
\section{ Quantum Cross Section for Electromagnetic Processes Induced
By Heavy Ions}

	We begin with a review of the results of Ref. \cite{10}, where the 
cross section
for nuclear excitation induced by the electromagnetic fields of a  passing 
heavy ion was derived in the first Born approximation. The relevant Feynman
diagram is shown in Fig. 1.
A {\it virtual} photon of momentum $q^\mu$ is
exchanged between the target and projectile nuclei, producing an excitation
in the target of energy $\omega_T$. (As a matter of convention, 
the nucleus that gets excited is considered to be  the target.) The cross 
section for simultaneously 
exciting both nuclei has been shown to be small\cite{11} at this 
level of approximation in the fine structure constant.
The lack of projectile
excitation and the large masses of the nuclei combine to determine 
both the
energy transfer and the component of the momentum transfer 
 along the direction of the projectile momentum. 
These are given, in the target rest frame, by
\begin{eqnarray}
q_0 &=& \omega_T\nonumber\\
{\bf q}^\parallel &=& \omega_T/\beta,
\end{eqnarray}
where $\beta$ is the projectile velocity. Our metric is such that
$q^2 \equiv q^2_0 - {\bf q}^2 $ . 

	In Ref. \cite{10}, the cross section was written in the compact form
\begin{eqnarray}
\sigma_{HI} &=& {2(Z_P\alpha)^2\over\beta^2}\int \,d\omega_T\,\rho(\omega_T)\int_{\omega_T/\beta}^
{\sqrt{(\omega_T/\beta)^2+{\bf q}_{max}^2}} \vert{\bf q}\vert d\vert {\bf q\vert} \vert F_P(q^2)\vert^2\left [ 
{1\over \gamma^2}{\vert F^T({\bf q})\vert^2\over 
(\omega_T^2-{\bf q}^2)
{\bf q}^2}
\right . \nonumber\\
 & &\left .\qquad\qquad\qquad\qquad\qquad\qquad -{\vert F^T({\bf
q})\vert^2 \over(\omega_T^2-{\bf q}^2)^2}+ {2\vert F^C({\bf q})\vert^2\over{\bf q}^4}\right ],
\end{eqnarray}
where $Z_P$ is the projectile's charge, $F_P(q^2)$ is the elastic form factor of the projectile, $F^T({\bf
q})$ and $F^C({\bf q})$ are the transverse and Coulomb form factors
of the target, $q_{max}$ is
a phenomenological cutoff on the transverse momentum transfer required to 
account for strong absorption effects, $\rho(\omega_T)$ is the
density of states in the target with excitation energy $\omega_T$,
and $\gamma=(1-\beta^2)^{-1/2}$.

	 From this expression, it is 
easy to see how the semi-classical limit is realized. As $\gamma$ becomes large,
the lower limit of the ${\bf q}$ integration approaches $\omega_T$.
When this happens, the behavior of the integral is dominated by the rapid 
variation of the poles at $q^2= 0$ in the photon
propagators. The nuclear densities and form factors vary much more slowly  
with $q^2$, and  are effectively frozen at their values for $q^2=0$. 
The third term in Eq.\ 2, which has no pole at $q^2=0$, is small compared
to the first two, which
grow logarithmically with $\gamma$. Thus, the cross section factorizes
neatly into a product of the 
same matrix elements  that appear in the
excitation cross section for real photons, times  an ``equivalent
photon number''. The latter is a function only of the 
$\omega_T$, $q_{max}$, and $\gamma$. 

	Outside of the large-$\gamma$ limit, there is no simple
factorization of the quantum cross section, Eq.\ (2),
so that there is no possibility for reconciliation with the semi-classical
expression  without further
approximations. The 
differences between the semi-classical approximation and the full
result are even more striking when the transverse momentum cutoff,
$q_{max}$, is removed to $\infty$. In this limit, the semi-classical
cross section diverges logarithmically, while the full expression
for the cross section remains finite.  There are three additional regulating 
factors in the full cross section, which tend to lower the cross section
even when $q_{max}$ is finite. The
first
factor arises because of the finite size of the projectile.
The
magnitude of the three-momentum 
transfer in the projectile's rest frame is given by $\sqrt{-q^2}\ge
\omega_T/\gamma\beta$, and the degree to which the protons in the 
projectile act coherently on the target is reduced at large $q$.
This produces
a cutoff governed by the size of projectile, $R_P$.  
The falloff of the target's excited state wave function at high momentum 
produces a corresponding 
second regulator governed by the target size, $R_T$. In the 
absence of both these effects, the cross section would still remain
finite as a result of the ${\bf q}^{-4}$ dependence of the integrals 
appearing in Eq.\ 2. This third factor
effectively cuts off the integrals at momentum scales of the order of 
$\omega_T$. If, in order to agree with semi-classical estimates, one chooses
$q_{max}\approx {1\over (R_T+R_P)}$, the situation becomes complicated,
as all the cutoffs are of comparable size. Without further study,
it impossible to determine which of these factors are most important in 
relation to  the measured cross section. 

\section{ Effects of Nuclear Structure}

	In this section, we examine the effects of  
the detailed form factors (or transition densities) on the cross sections
calculated with Eq.\ 2. These effects are naturally divided into projectile
and target structure, and we begin with the former. 

\subsection{ Projectile Structure}

In the semi-classical description, the projectile is assumed to be
a point charge, and only the long-range Coulomb field of the projectile
 generates the photon flux.
From Eq.\ 2, it
is apparent that the extended nature of the projectile enters the cross
section  through its elastic form factor, $F_P(q^2)$, which accounts
for the incoherence of the electromagnetic fields produced by  
spatially separate regions of the projectile.
 The size
of this effect, which tends to decrease the heavy-ion cross section, is 
governed by $(\omega_T R_P/\beta\gamma)^2$, where $R_P$ is the charge radius of 
the projectile. For light-target, heavy-projectile combinations, there is
no guarantee that this is small unless $\gamma$ is large.

In Fig. 2,  we demonstrate the effect of the finite size of the
projectile on the calculated cross section.
The ratio of the quantum to classical cross sections for a 20 MeV
dipole excitation of a mass 41 target by a $^{197}$Au projectile, as a 
function of projectile energy is shown. Here, and in all the calculations
to follow, we choose
$q_{max}=1/b_{min}$, where 
\begin{equation}
b_{min}= (1.34\, {\rm fm}) \left (A_P^{1/3}+A_T^{1/3}-0.75\cdot (A_P^{-1/3}
+A_T^{-1/3})\right )
\end{equation}
is a commonly used minimum-impact-parameter cutoff in 
semi-classical calculations\cite{3}\cite{12}. 
The four curves represent
the results of assuming either a point projectile(dot-dash curve), or
a projectile with mean-square charge radius of 
$\langle r^2 \rangle_{ch}^{1/2}=$ 5.4 fm and form factor parametrized by
\begin{eqnarray}
F_P(q^2) &=& {3j_1(x)\over x} \hspace{2.2in} \mbox{  (Bessel)}\nonumber\\
 &=& \exp(-x^2)  \hspace{1.85in}    \mbox{(Gaussian)}\nonumber\\
 &=& {1\over 1+x^2} \hspace{2in} \mbox{(Monopole)}\: ,
\end{eqnarray}
with $x=\sqrt{-q^2r^2/6}$, and $r$ the root-mean-square charge radius of
the projectile. For each curve we have assumed that target
transition densities are given by the Goldhaber-Teller model\cite{13}.

For low-energy projectiles ($\approx$ 50 MeV/nucleon), the calculated
cross section is
 sensitive to the finite size of the projectile, and is smaller by a factor 
of 2-3 than the point-projectile result. For relativistic projectiles,
the reduction in cross section is less dramatic, being about 7\% at $\gamma=2$
and decreasing as the projectile energy increases. Once the charge radius
is fixed, however, the resulting cross section is insensitive to the further
details of the form factor, except at the lowest projectile energies, where 
a variation $\approx $ 20\%  remains.
We conclude that the effect of  projectile
size is  non-negligible for many nuclear 
transitions, particularly when the projectile energy is low.
We note, however, that the effect  is smaller 
for lower-energy transitions, since the minimum $q^2$ of the virtual
photon varies as $\omega_T^2$. 

\subsection{Target Structure Effects}

	We now turn to the issue of target structure. In the classical
prescription, the target is assumed to respond to the electromagnetic
field of the passing projectile in exactly the same fashion as it would
to a real photon. Hence, for the majority of the transitions of interest,
the long-wavelength approximation should be valid. In the quantum case,
this is not guaranteed, since the momentum transfer is bounded from below by
$\omega_T/\beta > \omega_T $, so that higher moments of the transition 
density may play a larger role than they do for processes mediated by real
photons.  To calculate the quantum cross sections we used transition
form factors taken from two models. The first of these
represents a generalized collective
model for nuclear giant resonances\cite{13},
 which is motivated by the Goldhaber-Teller
(GT) model for the giant dipole resonance. The 
second set of transitions 
are the  shell-model form factors classified\cite{djm}
by their SU(3) symmetries.

In the generalized GT model,  the
transition densities are taken to be the gradient of a spherically
symmetric
ground-state density. These collective transition densities are then 
correct to leading order in the long-wavelength limit, and the form factors 
for a given  multipole $\ell$  are given by
\begin{eqnarray}
\vert F^C_\ell({\bf \vert q\vert})\vert^2&=& \ell\, C_{\ell}\, 
j_\ell^2({\vert\bf q\vert}R_T)\, , \nonumber\\
\vert F^T_\ell({\bf \vert q\vert})\vert^2&=& {(\ell+1)\over \ell}
{\omega_T^2\over \bf q^2}\vert F_\ell^C({\bf \vert q\vert})\vert^2\;\;\;
.
\end{eqnarray}
Here $R_T$ is the target charge  radius, and $C_\ell$ 
is a constant chosen so as to saturate the appropriate 
photonuclear sum rule for multipole $\ell$\cite{10}.

In the shell model, appropriate linear combinations of the single-particle 
transitions can 
provide descriptions of 
either giant resonance states or non-collective
states.  By examining each of these we can 
explore the sensitivity of the  cross section to 
a wide range of transitions, including the so called ``retarded''
transitions. (The retarded transitions are those that  
do not contribute to the real-photon cross section in
the leading order of the 
long-wavelength approximation.)
The separation  into unretarded 
and
retarded transitions can be achieved by classifying the shell-model
form factors by
their transformation properties under the SU(3)
symmetry\cite{djm,elliott,harvey} of the three
dimensional harmonic oscillator.  This classification scheme
has the advantage
of
allowing us to 
identify easily those form factors that dominate
the cross section in the long-wavelength limit.

Table I lists the expressions  for the dipole Coulomb and
transverse electric form factors classified under SU(3).
As discussed in the appendix,
these are  linear combinations of the usual $jj$-coupled
transitions.
Following Donnelly and Haxton\cite{DH}, the form factors are  
expressed in terms of a  
polynomial in $y$,
where  $y \equiv (b{\bf q}/2)^2$ and $b$ is the 
shell-model oscillator size parameter. 
For a dipole transition between oscillator orbits with $Q_1$ and $Q_2$ quanta
there are $(Q_1+Q_2+1)/2$ distinct form factors, and
these are
labelled with SU(3) quantum numbers ($\lambda,\mu)=(1,0), \cdots
(Q_1-1,Q_2-1),(Q_1,Q_2)$.
The maximum power of $y$ appearing in any form factor is determined
only
by the orbitals involved in the transitions and is equal to $(Q_1+Q_2)/2$.
The lowest power of $y$ 
is determined by $(\lambda,\mu)$
and is equal to $(\lambda+\mu)$/2.
Thus, the SU(3) scheme provides the required
linear combinations of the shell-model
form factors, separating them into an unretarded transition and
a set of transitions retarded to various orders in ${\bf q}^2$. 
In the long-wavelength approximation,
only the $(\lambda,\mu)=(1,0)$
form factor contributes in leading order, and it 
contains all the allowed B(E1)
strength. The
(1,0) form factor is 
 the shell-model equivalent of the Goldhaber-Teller giant resonance, 
and can be obtained by differentiating the ground-state density
distribution.
The $(\lambda,\mu)=(2,1)$ and higher SU(3) form factors
represent the retarded dipole transitions, and do not contribute to real
photon processes in the long-wavelength limit.

	In Figs. 3 and 4, we show the ratio of the quantum to semi-classical
cross sections for a 20 MeV E1 excitation of a mass 17 and mass 41 targets
by $^{197}$Au, using the Goldhaber-Teller
 and the $(\lambda,\mu$)=(1,0) and (2,1) 
shell-model  transition 
densities. The Goldhaber-Teller density and the (1,0) 
shell-model density
yield very similar
results. The quantum cross
section is enhanced relative to the semi-classical cross section
for $\gamma$ near unity, it 
is suppressed at moderate
$\gamma$, and returns slowly to the semi-classical result as 
$\gamma$ becomes large.  The cross-section ratio for the retarded (2,1) SU(3)
transition density
 is
markedly different from that for the (1,0) density, being larger at
low $\gamma$, and approaching the semiclassical result from above as
$\gamma$ becomes large. At low $\gamma$, the additional enhancement can
in part be 
 traced to the additional powers of 
$\vert {\bf q}\vert =\omega_T/\beta$ that
appear in the form factors, each leading to an enhancement of the cross
section by $1/\beta$ relative to the semi-classical result.
The shapes of the curves in Figs. 3 and 4 are determined essentially by 
the leading-order $\vert \bf q\vert$-dependence of the transition form
factors, and by the relative normalization of the Coulomb and transverse
form factors.

	While the dipole excitations provide the bulk of the relativistic
heavy-ion-induced 
electromagnetic cross section, quadrupole transitions have been
estimated\cite{9} to contribute significantly to the semi-classical
cross section at moderate projectile energies. To investigate the 
sensitivity of the E2 contribution to target structure effects, we 
once again compare the predictions of the Goldhaber-Teller and shell
models. For the E2 transition densities, it  proves useful to
classify the densities as representations of the SU(3) oscillator 
symmetry group, and these are listed in Table II.
As in the case of the dipole transitions, the SU(3) classification
scheme separates the quadrupole form factors into unretarded and
retarded linear combinations of the single-particle transitions.
There are two SU(3) form factors that contribute to the
E2 photon strength in the long-wavelength approximation. The first of these
transforms as $(\lambda,\mu)=(1,1)$ and  corresponds
to transitions within the same major shell, and the second transforms as
$(\lambda,\mu)=(2,0)$ and  corresponds
to transitions across two major shells. The (2,0) transition is 
the shell-model equivalent of the giant quadrupole resonance.
As can be seen from Table II, all other SU(3) quadrupole
transitions are of higher order in the long-wavelength limit.

In Fig. 5,
the ratio of the quantum to semi-classical cross sections are shown for
a 20 MeV E2 excitation of a
mass 41 target by $^{197}$Au, using the (2,0) and (2,2) shell-model form
factors, and the quadrupole transition density from the
Goldhaber-Teller model. Qualitatively, the results are  very similar to 
what was seen for the dipole cross section. The Goldhaber-Teller and
(2,0) transition densities, which have the same behavior in the 
long-wavelength limit, yield very similar results for the 
electromagnetic cross section. 
The retarded (2,2) transition, whose form factors
grow more rapidly at small $\vert{\bf q}\vert $, shows a more  
enhanced quantum to semi-classical ratio  
relative to that seen for the  (2,0) and Goldhaber-Teller densities. 
In all cases, the quantum E2 cross section is reduced dramatically from the
semi-classical result for all but the lowest projectile energies,
it is enhanced at very small $\gamma$, and returns to the
semi-classical result as $\gamma$ becomes large. 

	No other multipoles contribute measurably to the cross section
at relativistic  energies. 
Our  studies of the sensitivity of the
cross-section to the shape of the form factor show 
that only the leading-order $\vert {\bf q}\vert$-dependence and
relative normalization of the transverse and Coulomb transition
densities are necessary to provide an adequate description of the
heavy-ion-induced electromagnetic cross section.

\section{Simple Model for Transition Densities, Cross Sections}	

	In this section, we combine the results of the previous
sections to parametrize  the projectile and target
transition densities that incorporate the nuclear structure details
necessary to describe adequately the
cross section. Our goal  is to rewrite the cross section
in a form that allows us to extract an ``effective photon spectrum''
that can be used with real-photon cross section data. This will provide
an essentially
model-independent prediction for the electromagnetic cross sections
in peripheral heavy-ion
reaction, that incorporates the corrections
for the kinematic and finite-size effects described in the preceding
sections. 

	For the projectile, we have seen that the 
heavy-ion-induced electromagnetic cross section is sensitive only to
the rms charge radius, so we approximate the projectile form factor as
\begin{equation}
F_P(q^2)=1+{q^2R_P^2\over 6}+{\cal O}(q^4)\, ,
\end{equation}
with $R_P$ the rms charge radius of the projectile.
	
	For the target, we  restrict our attention to 
the ``unretarded'' transitions, which dominate the real-photon
cross sections in the long-wavelength limit. In the previous section,
we saw that the form of the heavy-ion cross section was largely determined
by the two factors: the leading-order dependence of the transition form factors
on the momentum transfer, and the relative normalization of the Coulomb
and transverse form factors. For unretarded transitions, this
normalization is determined by Siegert's theorem\cite{19}, which
builds in the constraints of angular momentum and current conservation.
We can express the quantum mechanical  Coulomb
excitation cross sections in terms of the real photoexcitation cross
section using the following long-wavelength approximation for the
transition form factors of multipolarity $\ell$:  
\begin{eqnarray}
 \vert F_\ell^T({\bf q})\vert^2 &=& {\omega_T\sigma_\gamma(\omega_T)\over
\pi\alpha\rho(\omega_T)}\times \Big ( {{\bf q}^2\over \omega_T^2}\Big
)^{\ell-1}\nonumber\\
 \vert F_\ell^C({\bf q})\vert^2 &=& {\omega_T\sigma_\gamma(\omega_T)\over
\pi\alpha\rho(\omega_T)}\times {\ell\over \ell+1}\Big ( {{\bf q}^2\over 
\omega_T^2} \Big )^\ell\, ,
 \end{eqnarray}
 Here
$\sigma_\gamma(\omega_T)$ is the cross section for exciting the target
with a real photon of energy $\omega_T$.

	Inserting these expressions into Eq.\ 2 allows the integrations
over the momentum of the virtual photon to be performed explicitly.
Since the transition densities are proportional to
$\sigma_\gamma(\omega_T)$, the cross section takes the form
\begin{equation}
\sigma_{HI}= \int d\omega_T \sigma_\gamma(\omega_T) n_{E\ell}(\omega_T)\, ,
\end{equation} 
where the effective photon spectrum, $n_{E\ell}(\omega_T)$, plays the
same role as the virtual photon spectrum in the classical calculation. 
For E1 transitions we find,
\begin{equation}
n_{E1}(\omega_T)=\frac{2 Z_P^2\alpha}{\pi\omega_T\beta^2}
\left[\ln\left({\Lambda\beta\gamma\over\omega_T}\right)
\left(1+{1\over 3}\left({\omega_T R_P\over\gamma}\right)^2\right) 
-{\beta^2\over 2} +{\omega_T^2\over 2\gamma^2\Lambda^2}-{\Lambda^2R_P^2\over 6}
\right],
\end{equation} 
where $\Lambda^2=q_{max}^2+({\omega_T\over\gamma\beta})^2$. 

For
E2 transitions the improved effective photon spectrum is given by, 
\begin{eqnarray}
n_{E2}(\omega_T)
&=&\frac{2 Z_P^2\alpha}{\pi\omega_T\beta^2}
\left[\ln
\left({\Lambda\beta\gamma\over\omega_T}\right)
\left(\beta^2+{1\over 3}\left({\omega_T R_P\over\gamma}\right)^2\right) 
-{\beta^2\over 2} +{\omega_T^2\over 2\gamma^2\Lambda^2} \nonumber\right. \\
& &\left. +{2\over 3}{\Lambda^2\over \omega_T^2}
\left(1-{\omega_T^2R_P^2\over 6} 
+{3\over 4}{\omega_T^2R_P^2\over\gamma^2}\right)
-{((\Lambda^2+{\omega_T^2\over\beta^2})^2
-({\omega_T\over\beta})^4)R_P^2\over 9\omega_T^2}
\right].
\end{eqnarray}

	To test our approximations we have recalculated the
cross-section ratios appearing in Figs. 2-5 and found that the 
simplified expressions appearing above  reproduce the
results of the full calculation.

\section{ Comparison with Data}

	In Ref. \cite{10}, a crude estimate of the heavy-ion-induced
electromagnetic cross section was made by assuming that all of the E1/E2
strength was concentrated in a discrete state at the peak of the 
giant resonance energy. While this
led to a cross section that was significantly lower than that obtained
with the corresponding semi-classical calculation, the results were still
larger than data.  When the finite width of the giant resonance is
taken into account, the
semi-classical prediction for the heavy-ion cross section is reduced.
A similar
reduction was expected in the fully quantum-mechanical cross section
in reference 10, but a quantitative comparison with data could not
be performed without resorting to unjustified assumptions 
concerning the effects of nuclear structure. The central result of the
present work
is that with a few simplifying  assumptions, it is now feasible to 
calculate the heavy-ion-induced cross section in exactly the same manner
as one does semi-classically. 

We focus our attention on the data of Hill et al. \cite{20} for 
single-neutron removal,
leaving a consideration of all the data  
for a later effort. 
The advantage of this particular
data set is that the cross section for each target is studied with several
projectiles, simplifying the search for systematic
effects. The total single-neutron-removal cross section, 
including a component produced by
strong interactions in grazing collisions, was measured for each
projectile-target combination. In order to 
extract the electromagnetic cross section, it
is necessary first to subtract the strong interaction contribution.
For light projectiles, the extracted electromagnetic cross section is
sensitive to the method used in estimating the strong contribution. Tables
III and IV list the data sets for the electromagnetic contribution to 
the single-neutron-removal cross section for $^{197}$Au and $^{59}$Co 
targets. The cross sections accounting for the strong interaction 
contribution  obtained from
the limiting fragmentation scheme of
Ref. \cite{22} and from the  Glauber estimate of  Ref. \cite{12}
are listed in
columns four and five of the tables, respectively. 

	Also shown in Tables III and IV are theoretical predictions
for the cross sections obtained with the semi-classical equivalent
photon flux and with the effective photon fluxes derived in the last
section. The cross section is given by
\begin{equation}
\sigma_{HI}=\int_{\omega_{thresh}}^{\omega_{max}} d\omega_T  
[ n_{E1}(\omega_T)\sigma_{E1}^\gamma(\omega_T)
+ n_{E2}(\omega_T)\sigma_{E2}^\gamma(\omega_T) ]\, ,
\end{equation}
where $\omega_{thresh}$ is the threshold for single-neutron removal from
the target (8(11) MeV for Au(Co)), $\omega_{max}$ is an upper limit for
the integration, taken to be 50 MeV, $n_{E\ell}$ are photon fluxes, and
$\sigma_{E\ell}^\gamma$ are the cross sections for single-neutron
removal by a real photon of the indicated multipolarity.  

To separate the E1 and E2 contributions to the total photo-neutron cross
section, we assumed that
the E2 cross section is
dominated by the isoscalar giant quadrupole resonance, and
is described by\cite{9}
\begin{equation}
\sigma_{E2}^\gamma (\omega) = {\sigma_{EWSR}\, \omega^2\over
1+(\omega^2-\omega_{GQR}^2)^2/\omega^2\Gamma^2} \, ,
\end{equation}
where
$\sigma_{EWSR} = f{0.22 ZA^{2/3}\over \pi\Gamma/2}$ $\mu$b-MeV$^{-1}$,
$\omega_{GQR}$ is the energy of the giant quadrupole resonance, $\Gamma$
is the resonance width, and $f$ is the fractional saturation of
the energy-weighted sum rule. Values for these parameters were taken from
Ref. \cite{9}.  

For convenience, the total cross sections
$(\sigma_{E1}^\gamma+\sigma_{E2}^\gamma)$ for single-neutron removal were
not taken directly from data, but were calculated from the parametrizations
of Berman\cite{21}. These parametrized fits overestimate the total
photo-neutron cross-sections, which in turn leads to an overestimate of
the calculated heavy-ion cross sections. 
The size of this  effect can be estimated by comparing the
semi-classical predictions obtained using the parametrized fits with
those of reference 9, where the photo-neutron data were used directly.
This indicates use of the parametrized fits leads 
to heavy-ion cross sections
that are larger by about $5\%$ for Au and by about 10$\%$
for Co.

The predictions for the Coulomb-excitation 
cross section, using both the
semi-classical and quantum expressions for the equivalent-photon fluxes,
are listed in Table III for the Au targets
and in Table IV for the Co target. 
The quantum calculation provides a significantly improved description of the
data, particularly if the Glauber picture is used to estimate the strong
interaction contribution to the cross section. 

\section{Summary}

	We have examined the sensitivity of  heavy-ion-induced 
electromagnetic cross sections to the structure of the
target and projectile nuclei. For typical  transitions,
we have shown that the cross section is only mildly dependent on the projectile
charge radius, and  more sensitive to the leading $\vert{\bf q}\vert$ 
dependence and the 
relative normalization of the target's Coulomb and transverse
form factors. 

	Using these results, we extracted a new equivalent-photon spectra
for E1 and E2 transitions, based on the fully-quantum-mechanical cross 
section derived in Ref. \cite{10}.
The new spectra may be used in the same fashion as their semi-classical 
counterparts to obtain model-independent predictions for
electromagnetic processes in relativistic-heavy-ion collisions. 
Moreover, the new 
spectra provide an explanation for some anomalously small measured
cross sections
in terms of single-photon 
exchange, and leave little room for more exotic multi-photon
mechanisms required to explain these cross 
sections in a semi-classical analysis\cite{22,23}. 

\vspace{10pt}	

\centerline{\bf Acknowledgements}

	The work of C.B. and J.F. was performed under the auspices of
the U.S. Department of Energy. One of us(C.B.) gratefully acknowledges
the support of the Iowa Space Grant Cooperative and the hospitality
of the Physics Department of the Univ. of Northern Iowa during the
final stages of this work. 

\newpage
 
\centerline{\bf Appendix}
In this appendix we discuss 
the shell-model transition form factors and their classification under
the SU(3) scheme. We also discuss the  use of the continuity equation
for the relative normalizations of the Coulomb and transverse form
factors. 

The Coulomb and the electric transverse
form factors  
for a transition of multipolarity J
are defined in terms of the nuclear charge and current transition
densities as:
\begin{equation}
F^C_J(q) = \mbox{\small $\sqrt{\frac{4\pi}{Z^2}
\frac{2J_f+1}{2J_i+1}}$}\int_0^\infty
\rho_J(r)j_J(qr)r^2dr
\nonumber
\end{equation}
\begin{equation}
F^E_J(q) =
 \mbox{\small $\sqrt{\frac{4\pi}{Z^2}\frac{J+1}{2J+1}}$} \int_0^\infty
\rho_{J,J-1}(r)j_{J-1}(qr)r^2dr +\mbox{\small $\sqrt{\frac{4\pi}{Z^2}
\frac{J}{2J+1}}$}
\int_0^\infty \rho_{J,J+1}(r)j_{J+1}(qr)r^2dr\;,
\end{equation}
The transition charge and current densities are defined in terms of the
reduced matrix elements of the charge and current operators as:
\begin{equation}
\rho_J(r)=\int <J_f\mid\mid \rho(\mbox{\boldmath $r$})Y_J(\hat{\mbox{\boldmath
$r$}})
\mid\mid J_i>\;d\hat{\mbox{\boldmath $r$}}
\nonumber
\end{equation}
\begin{equation}
\rho_{J,J'}(r) =
  \int <J_f \mid\mid \mbox{\boldmath $J(r)\, \cdot \, Y$}\!_{JJ'1}
(\hat{\mbox{\boldmath $r$}})\mid\mid J_i>d\hat{\mbox{\boldmath $r$}}\, .
\end{equation}

Donnelly and Haxton\cite{DH} have derived expressions  for the
single-particle matrix elements of the 
electromagnetic operators that can be used with
one-body density-matrix elements (OBDMEs) defined in jj-coupling. 
The form factors appropriate to SU(3) coupling are linear combinations
of these. They  can be obtained 
by expanding the SU(3) OBDMEs in terms of the jj-OBDMEs. 
For this, we write
\begin{eqnarray*}
<J_f T_f\mid\mid[a^+_{(Q_10)}\tilde{a}_{(0Q_2)}]^{(\lambda\mu)\kappa
(\Delta L \Delta S)\Delta J \Delta T}\mid\mid J_i T_i>
=\qquad\qquad\qquad\qquad\qquad
 & & \\ 
    \sum_{\ell_1\ell_2}(-)^{Q_2}<(Q_10)\ell_1(\overline{0Q_2})\ell_2
\|(\lambda\mu)\kappa L> 
\sum_{j_1j_2}\left(\begin{array}{ccc} \ell_1& \frac{1}{2}& j_1\\
                                       \ell_2& \frac{1}{2}& j_2\\
                                        \Delta L& \Delta S& \Delta J
                                         \end{array} \right)
<f\|(a^+_{j_1}\tilde{a}_{j_2})^{\Delta J \Delta T}\|i> \, ,& & 
\end{eqnarray*}
where the unitary 9-J symbol is that equal to the 9-J symbol of Brink and 
Satchler\cite{BS}
  multiplied by $\hat{j_1}\hat{j_2}\hat{\Delta L}\hat{\Delta S}$ with 
$\hat{j}=\sqrt{2j+1}$. The SU(3)$\subset$R(3) Clebsch-Gordon coefficient
is that of Draayer and Akiyama\cite{DA},
and the reduced matrix elements are  defined by Brink 
and Satchler. 

Tables I and II list the resulting SU(3) form factors for dipole and
quadrupole transitions that do not involve 
spin-flip, i.e., $\Delta S=0$. For the dipole form factors we
consider dipole transitions across one ($\Delta\hbar\omega_0=1)$ and
across three($\Delta\hbar\omega_0=3)$ oscillator shells, and in the case
of the quadrupole form factors we consider in-shell
($\Delta\hbar\omega_0=0)$ transitions  and transitions across two shells
($\Delta\hbar\omega_0=2$). As can be seen from the tables, the SU(3)
classification corresponds to a separation of the shell-model densities
into an unretarded transition and into a set of transitions that are retarded
to various orders in ${\bf q}^2$. 
The unretarded dipole form factor transforms as $(\lambda,\mu)=(1,0)$,
and contains all the dipole photon strength in the long-wavelength
approximation. For the quadrupole transitions there are two unretarded 
form factors corresponding to $\Delta\hbar\omega_0=0$ and
$\Delta\hbar\omega_0=2$ transitions, and these transform as (1,1) and
(2,0) under SU(3), respectively.

In determining the relative normalizations of the Coulomb and
transverse form factors it is important to ensure that the continuity
equation is satisfied.
The continuity equation relates the transition current
densities $\rho_{J,J-1}$ and
$\rho_{J,J+1}$ to the transition charge density $\rho_J$ by:
\begin{eqnarray*}
 \mbox{\small $\sqrt{\frac{J}{2J+1}}$}\; q
\int_0^{\infty}\rho_{J,J-1}(r)j_{J-1}
(qr)r^2dr
  & \\
=  &  \frac{\omega_T}{c} \int_0^\infty \rho_J(r) j_J(qr) r^2 dr +
\mbox{\small $\sqrt{\frac{J+1}{2J+1}}$}\; q \int_0^{\infty}
\rho_{J,J+1}(r) j_{J
+1}(qr)
r^2 dr \, .
\end{eqnarray*}
Thus, the transverse electric form factor can be expressed
in terms of the Coulomb form factor and
the current densities $\rho_{J,J+1}$ as:
\begin{equation}
F^E_J(q^2) =  \frac{\omega_T}{c\;q}\mbox{\small $\sqrt{\frac{J+1}{J}}$}F^C_J
+\mbox{\small
$\sqrt{\frac{2J+1}{J}}$}\int_0^\infty\rho_{J,J+1}(r)j_{J+1}(qr)r^2dr\; .
\end{equation}
Alternatively, one could eliminate the current density
$\rho_{J,J+1}(r)$, and express $F^E_J$ in terms of $F^C_J$ and
$\rho_{J,J-1}(r)$.

For the $(\lambda,\mu$)=(1,0) dipole and (1,1) and (2,2)  quadrupole
transitions the integral involving the transition current $\rho_{J,J+1}$
is identically zero. Thus, 
 the relative normalization of the Coulomb and electric
transverse form factors
are the same as for the
Goldhaber-Teller transitions (Eq.\ 5). For the retarded
transitions the situation is more complicated,
and the relation between $F^T_\ell$ and $F^C_\ell$ generally 
depends
on $(\lambda,\mu)$. 
For harmonic-oscillator wave functions of oscillator frequency $\omega_0$,
 the integral 
$\int^\infty_0\rho_{J,J+1 }\;j_{J+1}\;r^2dr$ is proportional to
$\frac{\omega_0}{c\;q}\int^\infty_0\rho_j\;j_J\;r^2dr$, and the constants of
proportionality are given in Tables I and II. Note that Eq.\ (17) implies
that only one of two distinct combinations of current terms can be written
in terms of the charge density\cite{19}. The remaining term is determined 
by different
physics, and in our model this term is distinguished by $\omega_0$ rather than
$\omega_T$.

\pagebreak
\begin{itemize}
\centerline{\underline{Table Captions}}
\item{Table I} Shell-model form factors for dipole transitions in the SU(3)
classification scheme. The form factors are expressed in terms of the 
variable $y=(b${\bf q}/2)$^2$, where $b$  is the oscillator parameter.
$\hbar\omega_T$ is the transition energy and $\hbar\omega_0$ is the
oscillator energy (i.e., $b=\sqrt{\frac{\hbar}{m\omega_0}}$).
These form factors correspond to linear combinations of the usual
jj-coupled transition form factors. As can be seen from the tables
the SU(3) classification separates
the transitions into an unretarded transition transforming as
$(\lambda,\mu)$=(1,0),
and into a set of transitions that are retarded to various orders in
${\bf q}^2$.

\item{Table II} Shell-model form factors for quadrupole transitions in
the SU(3) classification scheme.
\item{Table III} Comparison of the single-neutron-removal cross section
calculated semi-classically with the fully quantum-mechanical
photon spectrum derived in the text with data on $^{197}$Au targets.
\item{Table IV}  Comparison of the single-neutron removal cross section
calculated semi-classically with the fully quantum-mechanical
photon spectrum derived in the text with data on $^{59}$Co targets.
\end{itemize}

\pagebreak
\newcommand{\STRUT}{\rule{0in}{3ex}}
\begin{centering}
\begin{tabular}{|r|l|r|c|} \hline
\multicolumn{4}{|c|}{\STRUT Table I}\\
\multicolumn{4}{|c|}{Dipole Form Factors $\Delta J=1^-\, (\Delta L= 1\, \Delta
S=0)$} \\[1.0ex] \hline
\multicolumn{1}{|c|}{$Q_1\rightarrow Q_2$}\STRUT &\multicolumn{1}{|c|} {($\lambda,\mu)$}
 &\multicolumn{1}{|c|}{F$^{Coul}$}  &\multicolumn{1}{|c|}{ F$^T$/F$^{Coul}$} 
  \\[1.0ex] \hline
$\Delta\hbar\omega=1\hbar\omega$ \STRUT & & & \\
 $p\rightarrow s$ & (1,0) & $\sqrt{\frac{2}{3}} y^{1/2}e^{-y}$ &
$\frac{\sqrt{2} \omega_T}{c\:q}$ \\
 & & &  \\
 $sd\rightarrow p$ & 
(1,0)& $\sqrt{\frac{8}{3}} y^{1/2}(1 - \frac{1}{2}y)e^{-y}$ &
$\frac{\sqrt{2} \; \omega_T}{c\:q}$ \\
 & & &   \\
 & (2,1) & - $\sqrt{\frac{2}{15}} y^{3/2} e^{-y} $ &
$\frac{\sqrt{2} \; \omega_T}{c\:q} -\frac{4\sqrt{2} \; \omega_0}{c\:q}$\\ 
  & & & \\
 $pf\rightarrow sd$& (1,0) &$\sqrt{\frac{20}{3}} y^{1/2}(1 - y +
\frac{1}{5}y^2)e^{-y} $ 
&
$\frac{\sqrt{2} \; \omega_T}{c\:q}$  \\
 & & &   \\
 & (2,1) & $-\sqrt{\frac{4}{5}} y^{3/2}(1 - \frac{1}{3}y) e^{-y}$ & 
$\frac{\sqrt{2} \; \omega_T}{c\:q} -\frac{4\sqrt{2}  \; \omega_0}{c\:q}$ \\ 
  & & & \\
 & (3,2) & $-\frac{2\sqrt{2}}{3\sqrt{35}} y^{5/2} e^{-y} $&
$\frac{\sqrt{2}\;\omega_T}{c\:q}+\frac{2\sqrt{2}\;\omega_0}{c\:q}$\\
  & & & \\
$sdg\rightarrow pf$ & (1,0) &$ \sqrt{\frac{40}{3}} y^{1/2} (1-\frac{3}{2}y
+ \frac{3}{5}y^2 - \frac{1}{15} y^3)e^{-y}$ & 
$\frac{\sqrt{2}\omega_T}{c\:q}$\\ 
  & & & \\
  & & & \\
$\Delta\hbar\omega_0=3\hbar\omega_0$  & & &\\
$pf\rightarrow s$ & (3,0)& $ \sqrt{\frac{4}{135}} y^{3/2}e^{-y}$&
$\frac{\sqrt{2} \omega_T}{c\:q}
-\frac{2\sqrt{2}\omega_0}{c\:q}$ \\
 & & &  \\
$sdg\rightarrow p$ & (3,0) & 
$\sqrt{\frac{8}{5}}y^{3/2}(1-\frac{1}{3}y)e^{-y}$ &
$\frac{\sqrt{2}\omega_T}{c\:q}-\frac{2\sqrt{2}\omega_0}{c\:q} $
\\
  & & & \\
& (4,1)&$\frac{2}{3\sqrt{35}}y^{5/2}$ &$ \frac{\sqrt{2}\omega_T}{c\;q}
 -
\frac{8\sqrt{2}\omega_0}{c\:q}$  \\
 & &  &\\ \hline\hline
\end{tabular}
\end{centering}
\newpage
\begin{centering}
\begin{tabular}{|r|l|r|c|}\hline
\multicolumn{4}{|c|}{\STRUT Table II}\\
\multicolumn{4}{|c|}{Quadrupole Form Factors $\Delta J=2^+ (\Delta L=2
\Delta S=0)$} \\[1.0ex] \hline
\multicolumn{1}{|c|}{$Q_1\rightarrow Q_2$} \STRUT &\multicolumn{1}{|c|}{ ($\lambda,\mu)$}
 &\multicolumn{1}{|c|}{F$^{Coul}$} & \multicolumn{1}{|c|} {F$^T/F^{Coul}$}
  \\[1.0ex] \hline
$\Delta\hbar\omega_0=0$\STRUT  & & & \\
 $p\rightarrow p$ & (1,1) & $-\sqrt{\frac{8}{15}} y e^{-y}$ &
$\sqrt{\frac{3}{2}}\frac{\omega_T}{c\:q}$  \\
 & &  & \\
 $sd\rightarrow sd$& (1,1) &$\sqrt{\frac{8}{3}} y(1 -  \frac{2}{5}y)e^{-y} $
&$\sqrt{\frac{3}{2}}\frac{\omega_T}{c\:q} $   \\
& & &  \\
 &(2,2)$\kappa=0$ & $-\sqrt{\frac{8}{1365}} y^{2} e^{-y}$ 
&$\sqrt{\frac{3}{2}}\frac{\omega_T}{c\:q} +   
\frac{7\sqrt{6}\omega_0}{c\:q} $  
   \\
 & & & \\
 & (2,2)$\kappa=2$ &
 $\sqrt{\frac{8}{325}} y^2 e^{-y} $ &$\sqrt{\frac{3}{2}}\frac{\omega_T}{c\:q}
-  \frac{5\sqrt{2}}{\sqrt{3}}\frac{\omega_0}{c\:q}$ \\
  &  & & \\ 
$pf \rightarrow pf$& (1,1) &$-\sqrt{8} y(1 -  \frac{4}{5}y +
\frac{2}{15} y^2)e^{-y} $ & $\sqrt{\frac{3}{2}}\frac{\omega_T}{c\:q}$   \\
 & & &\\
 & (2,2)$\kappa=0$ & $-\sqrt{\frac{8}{195}} y^{2}( 1-\frac{2}{7}y)e^{-y}$
&$\sqrt{\frac{3}{2}}\frac{\omega_T}{c\:q}+\frac{7\sqrt{6}\omega_0}{c\:q}$ \\
 & & & \\
 & (2,2)$\kappa=2$ & $\frac{2\sqrt{14}}{5\sqrt{13}} y^2(1-\frac{2}{7}y) e^{-y} $
&$\sqrt{\frac{3}{2}}\frac{\omega_T}{c\:q}
-\frac{5\sqrt{2}}{\sqrt{3}}\frac{\omega_0}{c\:q}
$\\ 
 & & & \\
 & (3,3) &$\frac{8}{21\sqrt{15}} y^3 e^{-y}$ &$\sqrt{\frac{3}{2}}
\frac{\omega_T}{c\:q}

$  \\
  & & &  \\
$\Delta\hbar\omega_0=2$  & & & \\
$sd\rightarrow s$& (2,0)&$\sqrt{\frac{4}{15}} y e^{-y}$ &$\sqrt{\frac{3}{2}}
\frac{\omega_T}{c\:q}$  \\
 & & &   \\
$ pf\rightarrow p$& (2,0)&$\sqrt{\frac{4}{3}}
y(1-\frac{2}{5}y)e^{-y}$ &$\sqrt{\frac{3}{2}}\frac{\omega_T}{c\:q}$
\\
 & & & \\
& (3,1) & $-\frac{4}{5\sqrt{7}} y^2 e^{-y} $
 &$\sqrt{\frac{3}{2}}\frac{\omega_T}{c\:q} -
\frac{5\sqrt{6}}{3}\frac{\omega_0}{c\:q}$ 
 \\
 & & &   \\ \hline\hline
\end{tabular}
\end{centering}
\newpage
\begin{centering}
\begin{tabular}{|c||c|c||c|c|}\hline
\multicolumn{5}{|c|}{\STRUT Table III}\\[0.5ex] \hline
\multicolumn{1}{|c||}{Projectile(Energy/nucleon)}\STRUT & 
\multicolumn{4}{|c|} {$\sigma_{HI}$(mb)}\\[0.5ex] \hline
\STRUT {}& {Classical} & {Quantum} & {Expt.(Lim. Frag.)} & 
{Expt.(Glauber)} \\[0.5ex] \hline
\STRUT $^{12}$C(2.1 GeV) & 48 & 40 & 75$\pm$14 & 51$\pm$ 7 \\
& & & &\\
$^{20}$Ne(1.7 GeV) & 117 & 96 & 151$\pm$13 & 107$\pm$ 13 \\
& & & &\\
$^{20}$Ne(2.1 GeV) & 125 & 105 & 153$\pm$18 & 133$\pm$ 11 \\
& & & &\\
$^{40}$Ar(1.9 GeV) & 352 & 288 & 348$\pm$34 & 315$\pm$ 30 \\
& & & &\\
$^{56}$Fe(1.9 GeV) & 684 & 553 & 601$\pm$54 & 552$\pm$ 52 \\
& & & &\\
$^{86}$K(1.0 GeV) & 1001 & 762 & 820$\pm$62 & 793$\pm$ 62 \\
& & & &\\
$^{139}$La(1.26 GeV) & 2472 & 1886 & 1970$\pm$130 & 1952$\pm$ 130 \\
& & & &\\
$^{197}$Au(1.0 GeV) & 3967  & 2912 & 3077$\pm$200 & 3066$\pm$200  \\
& & & &\\
$^{209}$Bi(1.0 GeV/nuc) &4211  & 3154  &3244$\pm$205 & 3233$\pm$205  \\
& & & &\\
$^{238}$U(0.96 GeV/nuc) & 5024  & 3630 & 3160$\pm$230 & 3248$\pm$210  \\
& & & &\\ \hline 
\end{tabular}
\end{centering}
\newpage
\begin{centering}
\begin{tabular}{|c||c|c||c|c|}\hline
\multicolumn{5}{|c|}{\STRUT Table IV}\\[0.5ex] \hline
\multicolumn{1}{|c||}{Projectile(Energy/nucleon)}\STRUT & 
\multicolumn{4}{|c|} {$\sigma_{HI}$(mb)}\\[0.5ex] \hline
\STRUT {}& {Classical} & {Quantum} & {Expt.(Lim. Frag.)} & 
{Expt.(Glauber)} \\[0.5ex] \hline
$^{12}$C \STRUT(2.1 GeV) & 9 & 8 & 6$\pm$9 & -5$\pm$ 5 \\
& & & &\\
$^{20}$Ne(2.1 GeV) & 23 & 20 & 32$\pm$11 & 30$\pm$ 7 \\
& & & &\\
$^{56}$Fe(1.9 GeV) & 122 & 98 & 88$\pm$14 & 72$\pm$ 9 \\
& & & &\\
$^{139}$La(1.26 GeV) & 413 & 304 & 302$\pm$40 & 304$\pm$ 40 \\
& & & &\\ \hline 
\end{tabular}
\end{centering}
\newpage
\begin{figure}


\vspace*{1.4truein}             
\includegraphics{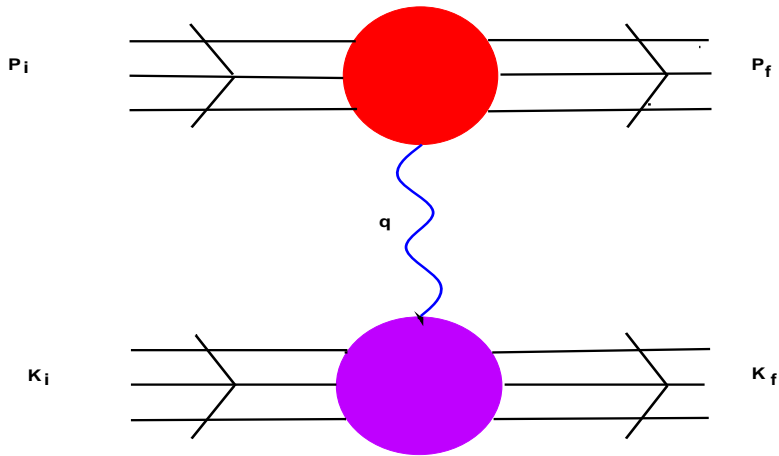}

\vspace*{2in}

\caption{ Feynman diagram for electromagnetic excitation in a
peripheral heavy-ion collision.}

\label{fig:ed}

\end{figure}
\newpage

\begin{figure}

\vspace*{13pt}


\vspace*{2.0truein}             
\includegraphics{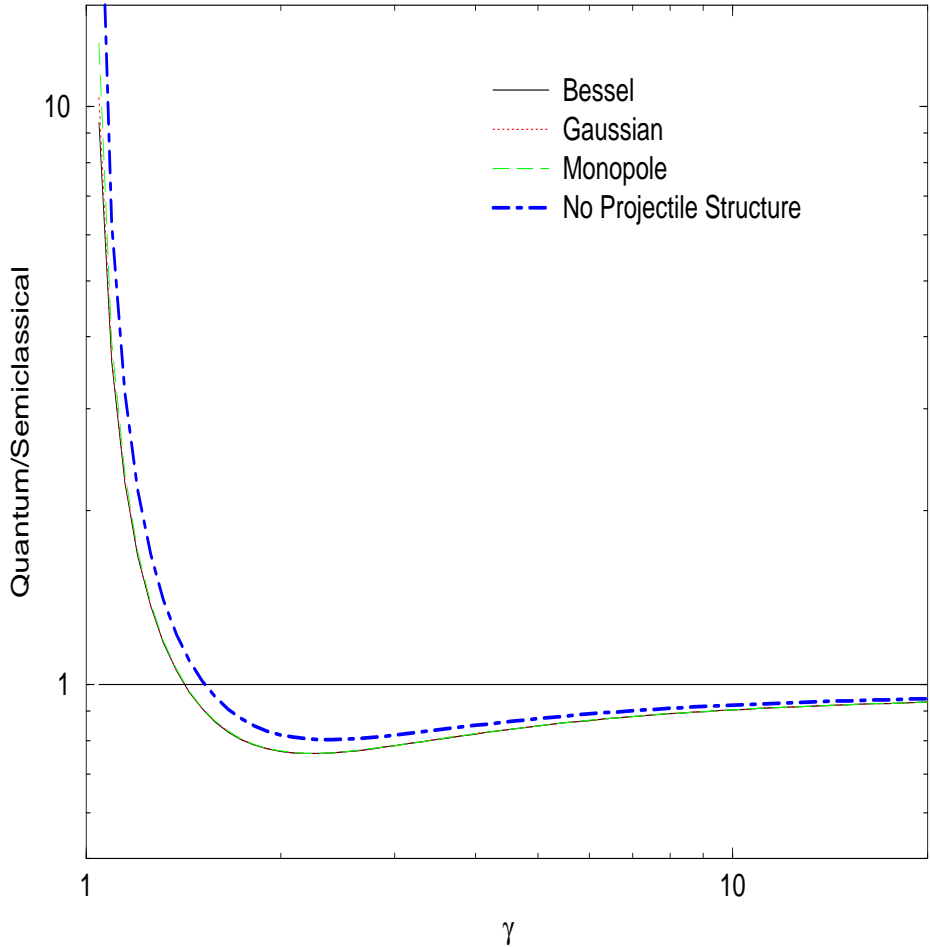}
\vspace*{4.in}
\caption{ The dependence of the Coulomb-excitation 
cross section on the finite size of the
projectile.
The figure shows the ratio of the full  quantum result
 to the standard 
semi-classical approximation for the cross section
 for a 20 MeV dipole excitation of $^{41}$Ca by a
$^{197}$Au ion. Results are shown for  a point projectile (dot-dash curve)
and for three different forms for $F_P(q^2)$ with the same charge radius.
For  low-projectile energies the predictions of the usual semi-classical 
approximation deviate significantly from the
full quantum results even for point-like projectiles. 
When the finite size of the projectile is included, the cross sections
shows sensitivity to the charge radius of the projectile, but
not to the detailed form of the projectile charge distribution.
For low-energy projectiles ($\approx$ 50 MeV/nucleon)
 the cross section is
reduced by a factor of 2-3 relative to the point-projectile result. At
$\gamma$=2 the calculated cross section is reduced by $7\%$ relative to
the point-projectile results, and the relative 
reduction
decreases as the projectile energy increases.}

\label{fig:projectile}

\end{figure}
\newpage
\begin{figure}

\vspace*{13pt}


\vspace*{2.0truein}             
\includegraphics{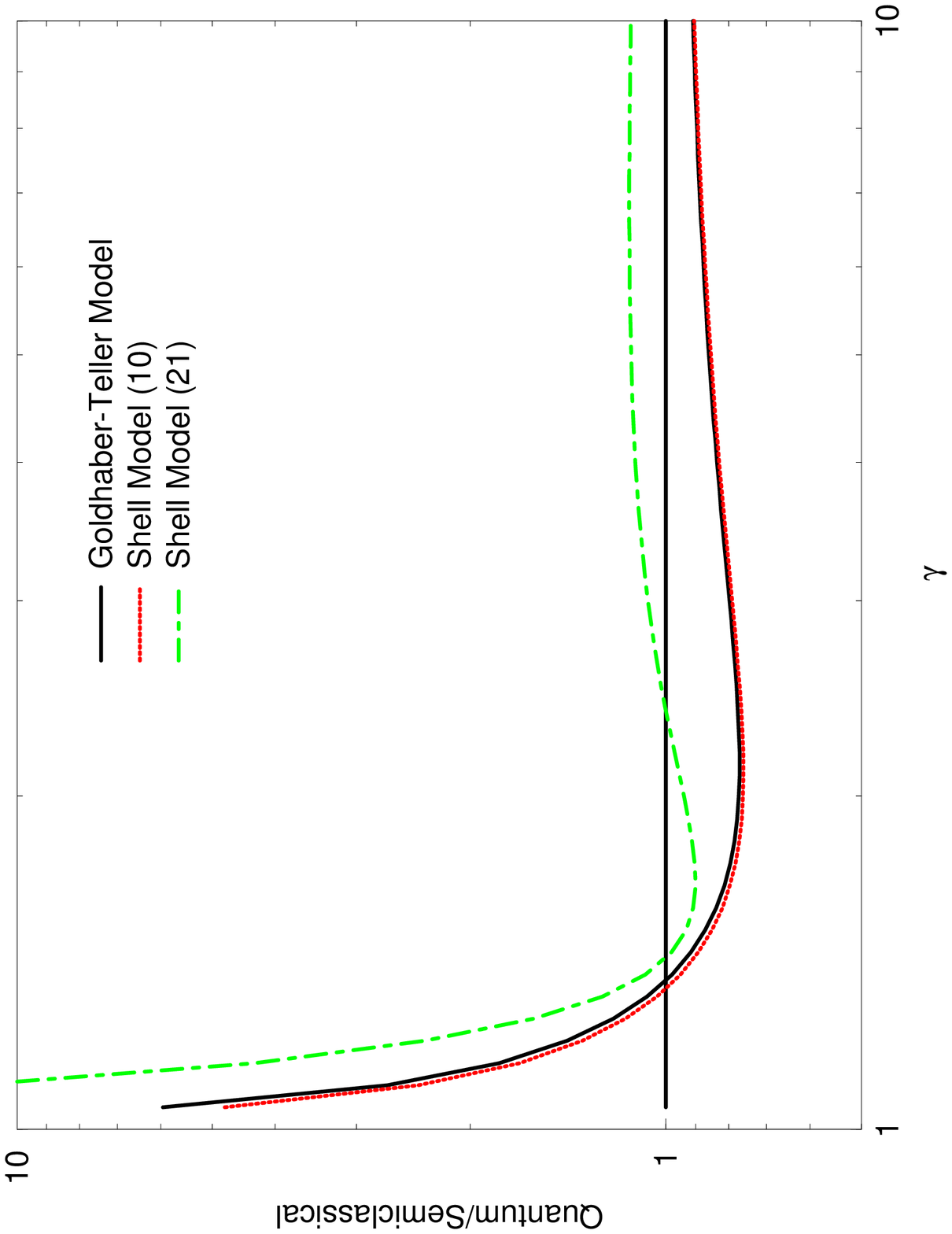}

\vspace*{4in}
\caption{ Ratio of the full quantum result to the standard 
semi-classical approximation for the 
cross section of a 20 MeV dipole excitation in $^{17}$O
by a
$^{197}$Au ion. The transition form factors for $^{17}$O
are described by either the Goldhaber-Teller Model, or by shell-model
densities transforming as $(1,0)$ or $(2,1)$ representations of SU(3).
The Goldhaber-Teller and the ($\lambda,\mu$)=(1,0) transition form
factors have the same leading-order dependence
in $\mid${\bf q}$\mid$ and their
calculated cross sections are very similar. The (2,1) transition form
factor does not contribute in leading order in the long-wavelength
approximation, and shows a markedly different cross-section ratio. In
all cases, the cross
sections are found to be sensitive to only the leading-order
$\mid${\bf q}$\mid$-dependence of the transition form factors and
to the relative normalization of the transverse and Coulomb form
factors. }

\label{fig:target17}

\end{figure}
\newpage
\begin{figure}

\vspace*{13pt}


\vspace*{2.0truein}             
\includegraphics{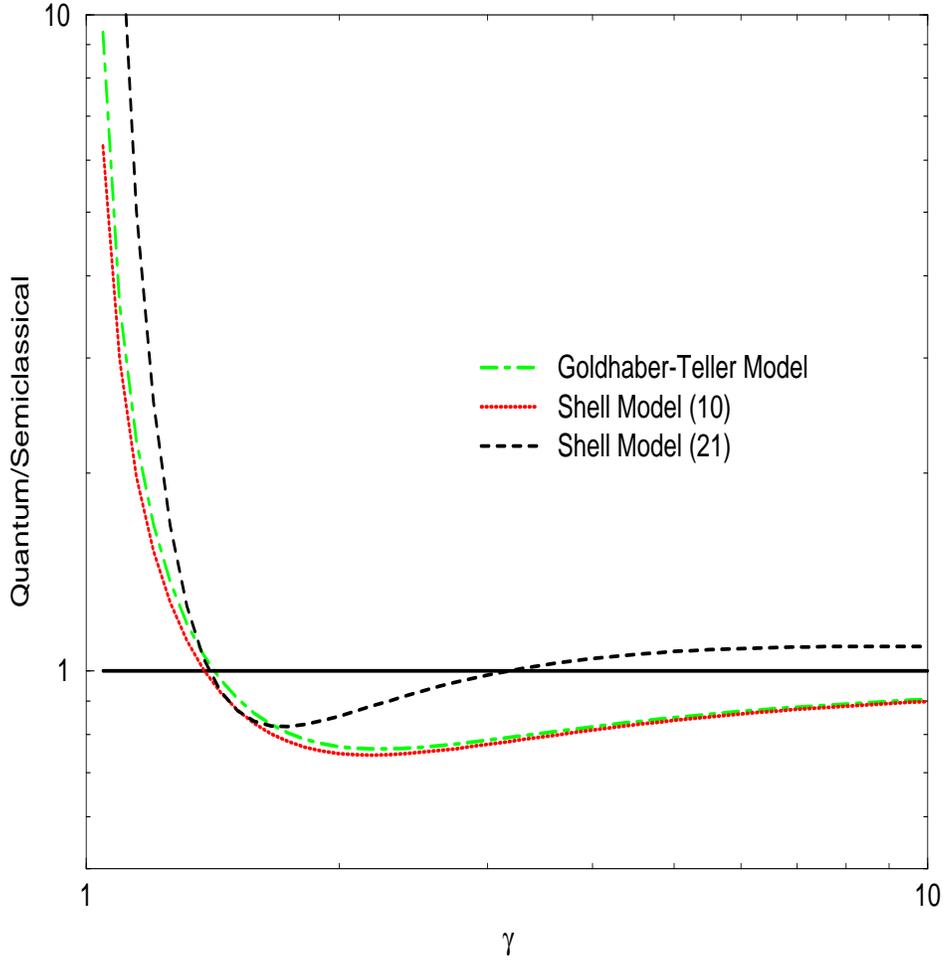}
\vspace*{4.in}

\caption{ Ratio of the full quantum result  to the
semi-classical approximation
for the cross section of a 20 MeV dipole excitation in $^{41}$Ca
by a
$^{197}$Au ion. The transition form factors of $^{41}$Ca
are described by either the Goldhaber-Teller Model, or by shell-model
densities transforming as $(1,0)$ or $(2,1)$ representations of SU(3).
As is the case of the calculations summarized in Fig. 3, 
the cross sections are  found to be sensitive
to only the leading $\mid${\bf q}$\mid$-dependence of the
transition form factors and to the relative normalization of the transverse
and Coulomb form factors.}

\label{fig:target41}

\end{figure}
\newpage
\begin{figure}

\vspace*{5pt}


\vspace*{2.0truein}             
\includegraphics{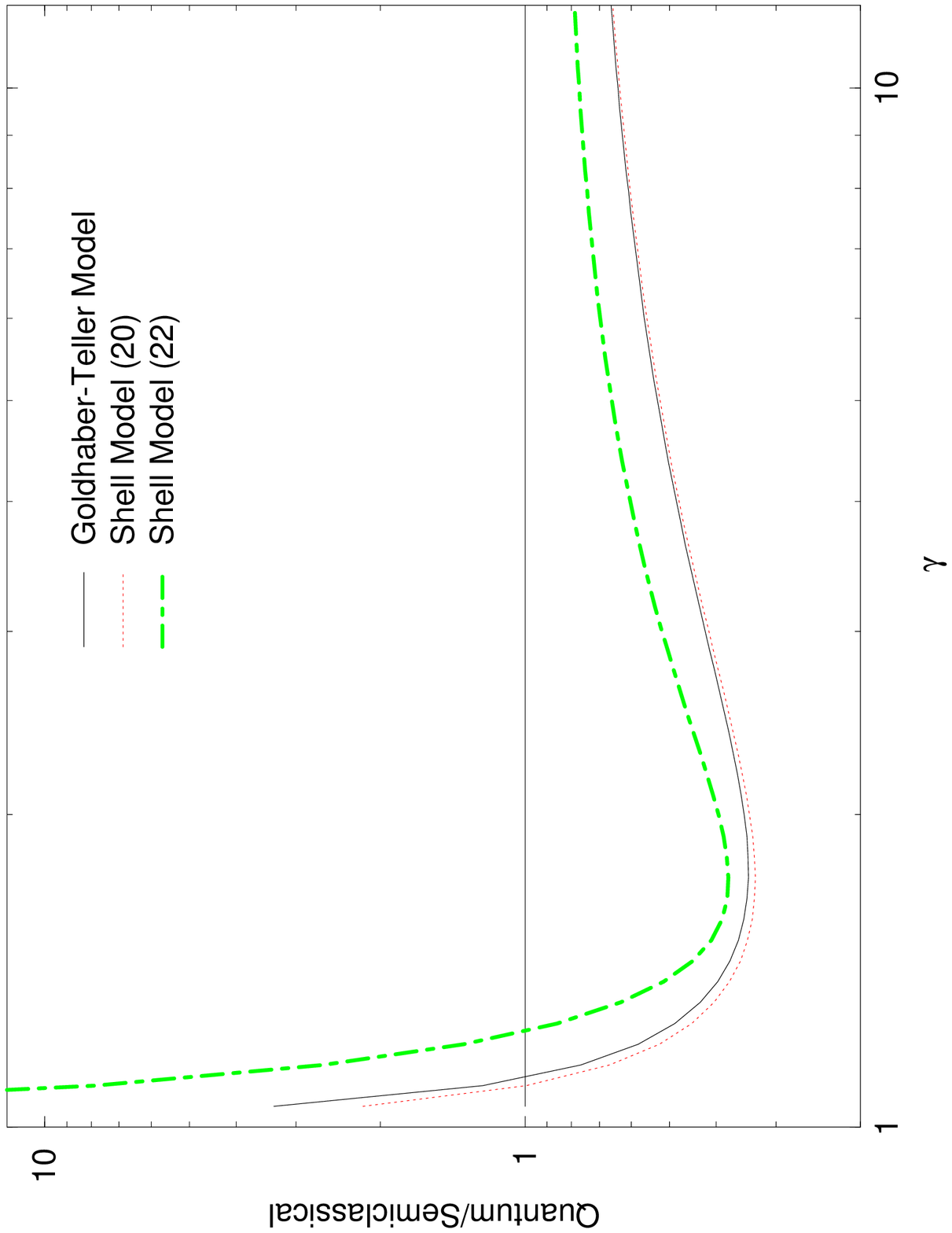}
\vspace*{4in}

\caption{ Ratio of the full quantum result to the
semi-classical approximation for the 
 cross section of  a 20 MeV quadrupole excitation in $^{41}$Ca
by a
$^{197}$Au ion.  The transition form factors for $^{41}$Ca
are described by either the Goldhaber-Teller Model, or by shell-model
quadrupole densities transforming
 as $(2,0)$ or $(2,2)$ representations of SU(3). The Goldhaber-Teller
transition and the  (2,0) SU(3) transition both represent
giant quadrupole transitions and 
involve the same leading 
 $\mid${\bf q}$\mid$-dependence
in the transition form factors.
The (2,2) SU(3) transition corresponds to a retarded transition that
does not contribute in leading order in the long-wavelength limit.
As in the case of the dipole transitions, the quadrupole cross sections
are found to be sensitive only to the leading term in the form factors
and to the relative normalization of the transverse and Coulomb form
factors.}

\label{fig:target41e2}

\end{figure}

\end{document}